\documentstyle[12pt]{article}
\renewcommand{\baselinestretch}{1.6}
\begin{document}
\newcommand{\be}{\begin{equation}}
\newcommand{\ee}{\end{equation}}
\newcommand{\al}{\alpha}
\newcommand{\G}{\Gamma}
\newcommand{\ve}{\varepsilon}
\title
{
\vspace*{-40mm}
\begin{flushright}
{\large \bf Preprint INR-0991/98\\[-5mm]
October 1998}\\[5mm]
\end{flushright}
\bf Top quark threshold  production in $\gamma\gamma$ collision
in the next-to-leading order.
}
\author{
  {\bf A.A.Penin and A.A.Pivovarov}\\
  {\small {\em Institute for Nuclear Research of the
  Russian Academy of Sciences,}}\\[-3mm]
  {\small {\em 60th October Anniversary
  Pr., 7a, Moscow 117312, Russia}}
       }

\date{}

\maketitle

\begin{abstract}
The total cross section of the top quark-antiquark 
pair production near 
threshold in $\gamma\gamma$ collision is computed analytically 
up to the next-to-leading order in perturbative and nonrelativistic
expansion for general photon 
helicity. The approximation includes the first order  corrections
in the strong coupling constant and  the heavy quark velocity
to the  nonrelativistic Coulomb approximation.
\\[2mm]
\end{abstract}

\thispagestyle{empty}
\newpage

\section{Introduction.}
Theoretical study of the top quark-antiquark 
pair production near the two-particle threshold 
is based on the key observation that the relatively large 
width $\G_t$ of the top quark 
serves 
as a natural infrared cutoff for long distance 
strong interaction effects \cite{FK}.
Quantitatively, the top quark width is mainly saturated by the decay channel 
$t\rightarrow Wb$ and is equal to $\G_t=1.43~{\rm GeV}$.
Near the production threshold the nonrelativistic
approximation is accurate and the relevant scale $\sqrt{\G_tm_t}=15.8~{\rm GeV}$,
$m_t$ being the top quark mass $m_t=175~{\rm GeV}$,
is much larger than $\Lambda_{\rm QCD}$
that makes perturbative 
QCD applicable for the theoretical description of the threshold
top quark physics \cite{FK,ttgg,ttee}. On the other hand 
the Next Linear Collider provides an opportunity 
of experimental study of high energy $\gamma\gamma$ interactions
which can be used for the top quark production. 
Using the laser backscattering method one can obtain 
$\gamma\gamma$ colliding beams with the energy and luminosity 
comparable to those in $e^+e^-$ collisions \cite{exp}. 
Because of high precision which can be achieved in such experiments
and the high accuracy of the theoretical description, the process
of top 
quark pair production 
in $\gamma\gamma$ collisions near the two-particle 
threshold has been recognized as a promising place 
for thorough investigation of quark interactions
and for determining the numerical values of 
the strong coupling constant $\al_s$,
the top quark mass $m_t$, and the top quark width $\G_t$. 
Moreover,
the strong dependence of the cross section on the photon 
helicities extends possibilities of studying
the details of the top quark threshold dynamics \cite{ttgg}. 

As is well known the ordinary finite order perturbation theory of QCD
breaks down in the threshold region of particle production
and the resummation 
of the Coulomb effects is necessary that can be
systematically done in the framework of nonrelativistic 
QCD (NRQCD) \cite{CasLep}. 
The threshold effects are treated exactly by using 
the nonrelativistic Schr\"odinger equation
with effective parameters.
This equation can be also obtained from the 
corresponding Bethe-Salpeter equation in the nonrelativistic limit.
Relativistic effects for the top quark are rather small 
and can be treated as corrections.
Note that the characteristic scale
of the Coulomb effects for the top production 
$\al_s m_t$ is comparable numerically with the 
cutoff scale $\sqrt{\G_tm_t}$ so
these effects are not suppressed by the top quark width.  
Quantitatively, Coulomb effects require to sum up power terms 
in the variable (up to color group factors)
$\al_s \sqrt{m_t/\G_t}\sim 1.1$ which is not small and 
the coresponding function is not accurately approximated by the finite
Taylor series.
The determination of the higher order strong coupling corrections
and 
relativistic corrections in this case 
requires 
the perturbative expansion to be performed 
near the Coulomb approximation rather than near the Green functions
of a free theory.
Technically, it makes the calculation rather involved.
Only recently an essential progress has been achieved 
in analytical evaluation of the next-to-leading order (NLO) 
and the next-to-next-to-leading order (NNLO)
corrections to the heavy quark vacuum polarization 
function near the threshold which enters the analysis
of the heavy quark threshold production in $e^+e^-$
annihilation \cite{KPP,Hoang,Mel,PP,MY,PP1}. The corrections
are found to be relatively large in the case of the top quark production
and very large in the case of the bottom quark threshold
sum rules. One can expect the similar effect 
in the $\gamma\gamma$ threshold production 
of heavy quarks.
Therefore, from phenomenological point of view,
the calculation of $\al_s$ and 
relativistic corrections to the corresponding cross section
is necessary for accurate quantitative study of the process. 
A part of the NLO corrections
(the one to the cross section
for the colliding photons of the same helicity)
has been studied in ref.~\cite{ttgg} using the 
numerical algorithm \cite{ttee}. The complete 
analytical result for the corrections is still absent.

In the present paper we formulate the relevant 
(NRQCD) framework for the calculation of the 
higher order corrections and obtain analytically   
the NLO corrections which include the 
first order corrections in the strong coupling constant 
and the heavy quark velocity to the nonrelativistic 
Coulomb approximation for general photon helicity.
The analytical result in the case of the 
opposite helicity colliding photons is 
of special importance because the available numerical 
methods are not applicable in this case \cite{ttgg}. 
We  evaluate the relativistic $\G_t\al_s$ corrections 
to the cross section for the opposite helicity colliding photons 
that cannot be found within the pure nonrelativistic approximation
and require an additional information on the relativistic structure of
the full theory. The procedure of matching to the full theory is used 
to recover the necessary parameters of effective NRQCD.
We present also the analytical NNLO result for the 
parameters of the resonances 
in the cross sections for $\gamma\gamma\rightarrow t\bar t$ and 
$e^+e^-\rightarrow t\bar t$ top quark threshold production.

The paper is organized as follows.
In the next Section we formulate the nonrelativistic 
approximation for the total cross section of top quark pair
production in 
$\gamma\gamma$ collisions. In 
Section~3 corrections to the nonrelativistic 
Green function and its derivative which enter
the expression for the cross section are considered.
In Section~4 we discuss the effects of nonvanishing top quark 
width. In Section~5 we present the NNLO analysis
of the resonance structure of the cross section. 
The last Section contains the results of numerical analysis
and our conclusions. 

\section{The nonrelativistic expansion for the cross
section.}
We fix our notation of kinematics
for the top quark pair
production in 
$\gamma\gamma$ collisions for working out
the nonrelativistic expansion for the cross
section as follows. 
The collision process is given by
\[
\gamma(q_1,\epsilon(q_1))\gamma(q_2,\epsilon(q_2))
\rightarrow t(p_1,\sigma_1) \bar t(p_2,\sigma_2)
\]
where $q_1,q_2$ are momenta of initial photons, $p_1,p_2$ are momenta
of the final state fermions
(top quark-antiquark pair), 
$q_1+q_2=p_1+p_2$, $(q_1+q_2)^2=s$ is the square of the 
total energy of two photon beams in the photons zero-momentum frame.
Here $\epsilon^\lambda(q_i)$ are the photon 
polarization vectors, $\sigma_{1,2}$ are spin variables
of the fermions.
After reducing the photon states the
amplitude of $\gamma\gamma\rightarrow t \bar t$ 
transition reads 
\be
{\cal M}(\gamma\gamma\rightarrow t\bar t)
=(eQ_t)^2\epsilon^{\mu}(q_1)\epsilon^{\nu}(q_2)
\langle t\bar t|T_{\mu\nu}|0\rangle 
\label{amp}
\ee
where $eQ_t$, $Q_t=2/3$, is the top quark electric charge.  
The quantity $T_{\mu\nu}$ in eq.~(\ref{amp})
is a Fourier transform of the $T$-product of two top quark
electromagnetic (vector) currents $J_\lambda=\bar t\gamma_\lambda t$
\be
T_{\mu\nu}(q) = i\int TJ_\mu(x/2)J_\nu(-x/2) e^{iqx}dx \ ,
\label{Tdef}
\ee
where 
$q=(q_1-q_2)/2$
is the difference between the phonon momenta. 

The $T$-product of the electromagnetic currents 
in eq.~(\ref{Tdef})
can be expanded into a series over the local
operators using a consistent expansion
in $x$ at small $x$ within the Wilson operator product
expansion \cite{pivpos}. 
Being inserted in eq.~(\ref{Tdef}) it converts into
a power series in  the variable $1/(m_t^2-q^2)$. Because  
$q^2=-s/4$ where $s$ is a total photon energy squared 
the expansion is not singular near the fermion-antifermion threshold
where $s\sim 4m_t^2$.
In fact, in the 
threshold region $q^2\sim -m_t^2$ and  
the corresponding expansion becomes an expansion in $1/m_t^2$.  
The leading term of the operator product
expansion (a contribution of the local operator of dimension three) 
reads \cite{pivpos}
\be
T^{(3)}_{\mu\nu}={2i\epsilon_{\mu\nu\alpha\beta}q^\alpha
\bar{t}\gamma^\beta \gamma_5 t \over m_t^2-q^2}\ .
\label{T3}
\ee
Next contributions to the transition amplitude 
are related to the dimension four local operators
of the operator product
expansion which have the following explicit form 
\be
T^{(4a)}_{\mu\nu}={i 
\bar t\gamma_{(\mu}\stackrel{\leftrightarrow}{D}_{\nu)} 
t 
\over m_t^2-q^2},
\label{T4a}
\ee
\be
T^{(4b)}_{\mu\nu}=
{2 q^\alpha m_t \bar t\sigma_{\mu\nu}
\stackrel{\leftrightarrow}{D}_\alpha t 
\over (m_t^2-q^2)^2},
\label{T4b}
\ee
\be
T^{(4c)}_{\mu\nu}=
{i 2 q^\alpha  \bar t q_{(\mu}\gamma_{\nu)}
\stackrel{\leftrightarrow}{D}_\alpha t
\over (m_t^2-q^2)^2}
\label{T4c}
\ee
where $D_\alpha$ is a covariant derivative containing 
the gluon field.
Note that operator~(\ref{T4c}) does not contribute
to the amplitude  because $q^\mu e_\mu(q_i)=0$.  
The operator product expansion 
is organized in such a way that near the two-particle threshold 
the matrix element of a higher dimension operator 
between the vacuum and $t\bar t$ state is suppressed 
by either the quark pair energy $E=\sqrt{s}-2m_t$ 
counted from the threshold or explicitly by powers of $\al_s$.
Thus, for the transition amplitude the operator product expansion 
converts into a simultaneous expansion
in $\al_s$ and 
in $\beta=\sqrt{1-4m_t^2/s}$,
the parameter of nonrelativistic near-threshold expansion. 
The operators in eqs.~(\ref{T3}--\ref{T4c}), however, 
are relativistic ones. Therefore, to obtain
the consistent nonrelativistic expansion of the amplitude
one has to expand these operators in a series over
the nonrelativistic operators in the framework of
NRQCD \cite{Bod}. The nonrelativistic 
expansion of the  operators entering eqs.~(\ref{T3},~\ref{T4a})  
which are necessary for the further analysis
up to a perturbative normalization factor reads 
\be 
\bar{t}\gamma^0\gamma_5 t = \varphi^\dagger\chi
+\ldots
\label{T3nonrel}
\ee
\be
i\bar t\gamma_{(i}\stackrel{\leftrightarrow}{D}_{j)}t 
=i\varphi^\dagger\sigma_{(i}\stackrel{\leftrightarrow}{D}_{j)}\chi +\ldots
\label{T4anonrel}
\ee
where $\varphi$ ($\chi$) is the nonrelativistic 
two component top (antitop) quark spinor, $\sigma_i$
is the Pauli matrix and ellipsis stands for
the higher dimension operators. Note that only the time component 
of the current in  eq.~(\ref{T3}) and space components 
of the tensor in   eq.~(\ref{T4a})  contribute in the leading order
of nonrelativistic expansion. 

Let us turn to the calculation of the cross section.
We deal with the normalized cross section
\[
R(s)={\sigma(\gamma\gamma\rightarrow t\bar t)\over
\sigma(e^+e^-\rightarrow\mu^+\mu^-)}
\]
where the standard lepton cross section
\[
\sigma(e^+e^-\rightarrow\mu^+\mu^-)={4\pi\al_{QED}^2\over 3s}
\]
is just a kinematic normalization factor.

In the case of the same helicity colliding photons
only the operator of dimension three contributes
to the amplitude up to NLO. 
Using the leading term of the nonrelativistic expansion 
for the quark current in eq.~(\ref{T3nonrel}) one arrives at the following
representation for the cross section
\cite{ttgg}
\be
R^{++}(E)=6Q_t^4N_cC_h^{++}(\al_s)\left({m_t^2\over 4\pi}\right)^{-1}
{\rm Im}G(0,0,k)
\label{Rpp}
\ee
where $G({\bf x},{\bf y},k)$ $(k^2=-m_tE)$ is the nonrelativistic
Green function which accounts for the soft (threshold) effects  and 
\[
C_h^{++}(\al_s)=1-\left(5-{\pi^2\over 4}\right){C_F\al_s(\mu_h)\over \pi}
\]
is the NLO perturbative coefficient matching correlators of relativistic 
and nonrelativistic  currents \cite{HarBr} with $\al_s$
being taken at ``hard'' normalization scale $\mu_h$.  
The higher dimension operators in the expansion of  eq.~(\ref{Tdef})
and in the nonrelativistic expansion~(\ref{T3nonrel})
lead to relative $O(\beta^2)$ corrections to the cross section.

For the opposite helicity colliding photons
the amplitude up to NLO is determined by the operator 
given in eq.~(\ref{T4a}).
By using the leading term of  eq.~(\ref{T4anonrel}) 
for the corresponding cross section 
one gets \cite{ttgg}
\be
R^{+-}(E)=8Q_t^4N_cC_h^{+-}(\al_s)\left({m_t^4\over 4\pi}\right)^{-1}
\partial_{\bf x}\partial_{\bf y}
{\rm Im}G({\bf x},{\bf y},k)|_{y=0,x=0}
\label{Rpm}
\ee
with the perturbative coefficient being \cite{BarKuh}
\[
C_h^{+-}(\al_s)=1-4{C_F\al_s(\mu_h)\over \pi}.
\]
Note that the leading contribution
to  $R^{+-}(E)$ comes from the quark pair with the orbital momentum $l=1$
because $l=0$ state cannot be produced by the opposite helicity
photons with the total angular momentum $J=2$  
while the dimension three operator~(\ref{T3}) interpolates only  
$l=0$ states and, therefore, does not contribute to $R^{+-}$ cross section.
The dimension four operator~(\ref{T4b})  does not contribute
to $R^{+-}$ in the leading order because it has  the parity $P=-1$
while the quark pair with the orbital momentum $l=1$ 
in the final state has the parity $P=1$.
In full analogy with the $R^{++}$ case the higher dimension operators 
lead only to relative $O(\beta^2)$ corrections to $R^{+-}$ cross section.

The nonrelativistic Green function appearing in eqs.~(\ref{Rpp},\ref{Rpm})
in the considered approximation
satisfies
the Schr{\"o}dinger equation
\be
\left(-{\partial_{\bf x}^2\over m_t}+V_C(x)+\Delta_1V(x)
+ {k^2\over m_t}\right)G({\bf x},{\bf y},k)=\delta({\bf x}-{\bf y})
\label{Schr}
\ee
where  $x=|{\bf x}|$, $V_C(x)=-C_F\al_s(\mu_s)/x$ is the  
Coulomb potential with $\mu_s$ being the soft normalization 
scale which determines the strong coupling constant. 
In the NLO 
one has to take into account the
first order perturbative QCD corrections to the Coulomb
potential $\Delta_1V$ which is of the following form \cite{Fish}  
\be
\Delta_1V(x)={\al_s\over 4\pi}V_C(x)(C_0^1+C_1^1\ln(x\mu_s))=
{\al_s\over 4\pi}V_C(x)C_1^1\ln(x\mu_1)
\label{potcorr}
\ee
where
\[
C_0^1=a_1+2\beta_0\gamma_E,\qquad C_1^1=2\beta_0,
\]
$$
a_1={31\over 9}C_A-{20\over 9}T_Fn_f ,\qquad 
\mu_1=\mu_s {\rm exp}\left({C_0^1/C_1^1}\right).
$$
The color symmetry
$SU(3)$ group invariants for QCD are $C_A=3$, $C_F=4/3$, $T_F=1/2$,
$n_f=5$ is the number of light fermion flavors, and
$\beta_0=11C_A/3-4T_Fn_f/3$ is the first coefficient of $\beta$-function.
Here $\gamma_E=0.577216\ldots$ is the Euler constant.

Note that up to an overall normalization factor the expression 
for $R^{++}$ up to NLO coincides with the normalized cross section
of the top quark threshold production in $e^+e^-$
annihilation.  

In zero order in $\al_s$ eqs.~(\ref{Rpp},\ref{Rpm})
are reduced to Born approximation for corresponding cross sections
and contain only continuous 
spectrum starting from the two-particle threshold
\[
R^{++}(\beta)=6Q_t^4N_c\beta\theta(s-4m_t^2)+O(\beta^3),
\]
\[
R^{+-}(\beta)=8Q_t^4N_c\beta^3\theta(s-4m_t^2)+O(\beta^5).
\]
The full relativistic result for the cross sections 
$R^{++}(\beta)$ and $R^{+-}(\beta)$ in tree approximation
which sums up all powers of $\beta$ was obtained in 
ref. \cite{INOK}. At the one loop level the analytical 
result has been recently obtained as a power expansion in $\beta$
up to $\beta^{10}$
order \cite{KM}. 

\section{The nonrelativistic Green function beyond the leading
order.}
In this section we describe 
the technique for computing
the nonrelativistic Green function beyond the leading
order. The corrections are found within the perturbation theory around 
the Coulomb solution as a leading order approximation relevant for
computation of the cross section near the threshold.
The developed technique is applicable for computing corrections to 
the nonrelativistic Green function with any orbital momentum $l$ 
and is a direct generalization of the method suggested in
\cite{KPP,PP,PP1}.

The nonrelativistic Green function has the standard  
partial wave decomposition  
\be
G({\bf x},{\bf y},k)=\sum^\infty_{l=0}(2l+1)
(xy)^lP_l({\bf xy}/xy)G_l(x,y,k)
\label{sph}
\ee
where $P_l(z)$ is a Legendre polynomial.
The partial waves of the Green function of the pure 
Coulomb Schr{\"o}dinger equation 
$G^C({\bf x},{\bf y},k)$ are
\be
G^C_l(x,y,k)={m_tk\over 2\pi}(2k)^{2l}e^{-k(x+y)}
\sum_{m=0}^\infty {L_m^{2l+1}(2kx) L_m^{2l+1}(2ky)m!\over
(m+l+1-\nu)(m+2l+1)!}
\label{rep1}
\ee
where $\nu=\lambda / k$, $\lambda =\alpha_sC_Fm_t/2$, 
 $L^\al_m(z)$ is a Laguerre  polynomial
\[
L_m^\al(z)={e^zz^{-\al}\over m!}\left({d\over dz}\right)^m
(e^{-z}z^{m+\al})
\]
and $\al_s$ is taken at the soft scale $\mu_s$.
At $y=0$ the partial waves take  the form 
\be
G^C_l(x,0,k)=
{m_tk\over 2\pi}(2k)^{2l}e^{-kx}\Gamma(l+1-\nu)U(l+1-\nu,2l+2,2kx)
\label{rep}
\ee
where $U(a,b,z)$ is the  confluent hypergeometric function.

Only $l=0$ component of the Green function contributes to  
its value at the origin 
\[
G(0,0,k)=G_0(0,0,k).
\]
For the Coulomb Green function we find an explicit expression in the form
\[
G^C_0(x,0,k)|_{x\rightarrow 0}=
{m_t\over 4\pi}\left({1\over x}-
2\lambda\ln\left({2x\mu_f}\right)
-2\lambda\left({k\over 2\lambda}+
\ln\left({k\over \mu_f}\right)
\right.\right.
\]
\be
\left.\left.
+2\gamma_E-1+
\Psi_1\left(1- \nu\right)\right)\right)
\label{G0}
\ee
where $\Psi_n(z)=d^n\ln{\Gamma(z)}/dz^n$ and
$\Gamma(z)$ is the Euler $\Gamma$-function.
In the short distance limit $x\rightarrow 0$
the Coulomb Green function $G^C({\bf x},0,k)$ has $1/x$ and $\ln x$
divergent terms. These terms, however, are energy independent 
and do not contribute
to the cross section. Hence they can be subtracted. 
In eq.~(\ref{G0}) $\mu_f$ is an auxiliary parameter.  
Though an accurate definition of this parameter
is important in the NNLO analysis \cite{Hoang,MY} it does not 
enter the cross section of top production for the stable top quark, 
or in zero width approximation.  
 
The derivative of the Green function at the origin
is saturated with its $l=1$ component  
\[
\partial_{\bf x}\partial_{\bf y}
G({\bf x},{\bf y},k)|_{y=0,x=0}=
9G_1(0,0,k) 
\]
and  for the Coulomb Green function we have
\[
G^C_1(0,0,k)=
{m_t\over 36\pi}
\left({3\over x^3}+{3\lambda\over x^2}+
{6\lambda^2-3k^2\over 2x}+2\lambda(k^2-\lambda^2)\ln(2x\mu_f)
\right.
\]
\be
\left.
+\lambda\left(2(k^2-\lambda^2)\left({k\over 2\lambda}+
\ln\left({k\over \mu_f}\right)
+2\gamma_E-{11\over 6}+
\Psi_1\left(1- \nu\right)\right)+{k^2\over 2}\right)\right).
\label{dG0}
\ee
One can immediately see that this expression has 
no lowest pole present in eq.~(\ref{G0}).
In the short distance limit $x\rightarrow 0$
the derivative of 
Coulomb Green function 
(or partial wave with $l=1$) 
has $1/x^n$ $(n=1,2,3)$ 
and $\ln x$ singularities. In contrast to Coulomb 
Green function itself 
some of the divergent terms now
are $k$ dependent but also do not contribute
to the cross section for $\G_t=0$ because they have no discontinuity 
across the physical cut in the complex energy plane in the zero-width
approximation.
The case of  
non-zero top quark width needs more detailed analysis given
in the next section.

The solution to eq.~(\ref{Schr}) with the logarithmic
correction to the Coulomb potential can be found within the 
standard nonrelativistic perturbation theory
around the Coulomb Green function taken as a leading order approximation
\[
G({\bf x},{\bf y},k)=G^C({\bf x},{\bf y},k)+\Delta_1G({\bf x},{\bf y},k),
\]
\be
\Delta_1G({\bf x},{\bf y},k)=-\int G^C({\bf x},{\bf z},k)
\Delta_1V(z)G^C({\bf z},{\bf y},k)d{\bf z}\ .
\label{gfcorr}
\ee
The calculation goes along the lines developed  
in refs.~\cite{KPP,PP,PP1}\footnote{The correction to 
the Green function at the origin 
has been also studied using the numerical algorithm 
in refs.~\cite{ttgg,ttee,Hoang,Mel}.}. 
Only $l=0$ component of eq.~(\ref{sph}) is necessary
for the calculation of the correction to the Green 
function at the origin. The integral in eq.~(\ref{gfcorr})  
diverges at $x,~y= 0$. This divergence is induced by the singular $1/x$
term in eq.~(\ref{G0}) which is related to the singular  behavior
of the free $(\al_s=0)$ Green function $G^F({\bf x},{\bf y},k)$
in the limit  $y=0$, $x\rightarrow 0$. 
The divergent part can be separated in the following  way.
Eq.~(\ref{gfcorr}) for $x,y = 0$ can be written in the form
\[
\Delta_1G(0,0,k)=\Delta_1G_0(0,0,k)=-\left(\int 
(G_0^C(0,x,k)-G_0^F(0,x,k))^2
\Delta_1V(x)
d{\bf x}\right.
\]
\be
\left.
+2\int 
(G_0^C(0, x,k)-G_0^F(0,x,k))G_0^F(x,0,k)
\Delta_1V(x)d{\bf x}+
\int G_0^F(0,x,k)^2\Delta_1V(x)
d{\bf x}\right).
\label{G1a}
\ee
By using representation~(\ref{rep}) for $G_0^C(x,y,k)$
and the same expression with $\al_s=0$ for $G_0^F(x,y,k)$  
one gets
\[
\Delta_1G_0(0,0,k)=-\left({m_tk\over 2\pi}\right)^2\left(
\sum_{m,n=0}^\infty F(m)F(n)\int e^{-2kx}
L_m^1(2kx)L_n^1(2kx)\Delta_1V(x)d{\bf x}
\right.
\]
\be
\left.
+2\sum_{m=0}^\infty F(m)\int
{e^{-2kx}L_m^1(2kx)\Delta_1V(x)\over 2kx}d{\bf x}
+\int {e^{-2kx}\Delta_1V(x)\over (2kx)^2}d{\bf x}\right)
\label{G1b}
\ee
where
\[
F(m)={\nu\over (m+1)\left(m+1-\nu\right)}
\]
and we used the properties of the  Laguerre  polynomial
\[
L_m^\al(0)={\Gamma(m+\al+1)\over \Gamma(\al+1)\Gamma(m+1)}\,,\quad 
\sum_{m=0}^\infty{L_m^\al(z)\over m+\al}=z^{-\al}\Gamma(\al).
\]
Only the last term is divergent in eq.~(\ref{G1b}).
After regularization the divergent integral in that term 
becomes
\[
\int_0^\infty {e^{-2kx}\ln(\mu x)\over x}dx=
-\gamma_EL(k)+{1\over 2}L(k)^2+\ldots 
\]
where $L(k)=-\ln(2k/\mu_1)$ and
ellipsis stands for the divergent part.
The divergent part being $k$ independent
does not contribute to the spectral density.
Two finite integrals in eq.~(\ref{G1a}) are
\[
\int_0^\infty e^{-z}L^1_n(z)L^1_m(z)\ln(z)zdz=
\left\{
\begin{array}{c}
(m+1)\Psi_1(m+2), \quad m=n \\
{\displaystyle -{n+1\over m-n}}, \quad m>n \\
\end{array}
\right.
\]
\be
\int_0^\infty e^{-z}L^1_m(z)\ln(z)dz= -2\gamma_E-\Psi_1(m+1).
\label{integ1}
\ee
To compute the first integral we rewrite it in the form
\be
\left.
\int_0^\infty e^{-z}L^1_n(z)L^1_m(z)\ln(z)zdz=
{d\over d\ve}\left(\int e^{-z}L^1_n(z)L^1_m(z)
z^{1+\ve}dz
\right)\right|_{\ve =0}.
\label{integ2}
\ee
By using the relations
\[
L^\beta_m(z)=\sum_{n=0}^m{\Gamma(\beta-\al +n)\over\Gamma(\beta-
\al)\Gamma(n+1)}L_{m-n}^\al(z),
\]
\[
\int_0^\infty e^{-z}L^\al_n(z)L^\al_m(z)z^\al dz=
\delta_{mn}{\Gamma(m+\al +1)\over\Gamma(m+1)}
\]
for $\beta=1$, $\al=1+\ve$ the integration in
the right hand side of eq.~(\ref{integ2}) can be performed
analytically. Then taking the derivative in $\ve$ at
$\ve=0$ we get the first line of eq.~(\ref{integ1}).
The second integral in eq.~(\ref{integ1}) can be computed
using the same technique.
Thus the final result for the correction is
$$
\Delta_1G_0(0,0,k)={\al_s\beta_0\over 2\pi}{\lambda m_t\over 2\pi}\left(
\sum_{m=0}^\infty F(m)^2(m+1)
\left(L(k)+\Psi_1(m+2)\right)-2\sum_{m=1}^\infty\sum_{n=0}^{m-1}
F(m)
\right.
$$
\be
\left. \times F(n){n+1\over m-n} +2\sum_{m=0}^\infty F(m)
\left(L(k) - 2\gamma_E-\Psi_1(m+1)\right)
-\gamma_E L(k)+{1\over 2}L(k)^2
\right).
\label{G1}
\ee
Note that the sum in the finite limits
in this equation can be reexpressed through
some combination of $\psi$ functions of different arguments.
This however is not important for numerical analysis 
at present when the actual calculation is performed with
computers. 

The calculation of the correction to the 
derivative of the Green function at the origin (or partial wave 
with $l=1$) 
is a bit more cumbersome. 
The corresponding expression reads
\be
\partial_{\bf x}\partial_{\bf y}\Delta_1G({\bf x},{\bf y},k)|_{x,y=0}
=9\Delta_1G_1(0,0,k) =-9\int 
G_1^C(0,x,k)^2
\Delta_1V(x)x^2d{\bf x}.
\label{dG1a}
\ee
Eq.~(\ref{dG0}) contains $1/x^n$ $(n=1,~2)$ singular terms 
with the coefficients depending on $\al_s$ which lead to
the divergence of the integral in eq.~(\ref{dG1a}). 
Hence this is not enough
to subtract just the free Green function from the Coulomb one
to separate the divergent part  of eq.~(\ref{dG1a}) 
as has been done in the analysis of the correction to the  
Green function at the origin. 
The analog of eq.~(\ref{G1a}) now reads
\[
\Delta_1G_1(0,0,k)
=-\left(\int 
(G_1^C(x,0,k)-D(x,k))^2
\Delta_1V(x)
x^2d{\bf x}\right.
\]
\be
\left.
+2\int 
(G_1^C(x,0,k)-D(x,k))D(x,k)
\Delta_1V(x)x^2d{\bf x}+
\int D(x,k)^2\Delta_1V(x)
x^2d{\bf x}\right)
\label{dG1b}
\ee
where
\[
D(x,k)={m_tk^3\over 3\pi}e^{-kx}
H(2kx),
\]
\be
H(z)={2\over z^3}+{1+\nu\over z^2}+
{\nu(1+\nu)\over 2z}.
\label{D}
\ee
By using the relation
\[
H(z)=\sum_{m=0}^\infty L_m^3(z)
\left({\nu(\nu+1)\over 2(m+1)}+{1-\nu^2\over m+2}+
{\nu(\nu-1)\over 2(m+3)}\right)
\]
eq.~(\ref{dG1b}) can be transformed into
\[
\Delta_1G_1(0,0,k)
=-\left({m_tk^3\over 3\pi}\right)^2\left(
\sum_{m,n=0}^\infty \tilde F(m)\tilde F(n)\int e^{-2kx}
L_m^3(2kx)L_n^3(2kx)\Delta_1V(x)x^2d{\bf x}
\right.
\]
\be
\left.
+2\sum_{m=0}^\infty \tilde F(m)\int
e^{-2kx}L_m^3(2kx)H(2kx)\Delta_1V(x)
x^2d{\bf x}
+\int e^{-2kx}H(2kx)^2\Delta_1V(x)x^2d{\bf x}
\right)
\label{dG1c}
\ee
where
\[
\tilde F(m)={\nu(\nu^2-1)\over (m+2-\nu)(m+1)(m+2)(m+3)}
\]
and the divergence is contained in the last term.
After a regularization this term  can be directly
computed with the result
\[ 
I(k)=
-\left({C_F\beta_0\al_s^2\over 2\pi}\right)^{-1}
{4k^4\over \pi}\int e^{-2kx}H(2kx)^2\Delta_1V(x)x^2d{\bf x}
=-{(\gamma_E-1)^2\over 2}-{\pi^2\over 12}
\]
\be
-(4-3\gamma_E)\nu +{1-9\gamma_E+6\gamma_E^2+\pi^2\over 4}\nu^2+
{1-3\gamma_E\over 2}\nu^3+{1-\gamma_E\over 4}\nu^4
\label{Dinteg}
\ee
\[
+\left(\gamma_E-1-3\nu+{9-12\gamma_E\over 4}\nu^2
+{3\over 2}\nu^3+{1\over 4}\nu^4\right)L(k)
+\left(-{1\over 2}+{3\over 2}\nu^2\right)L(k)^2
+\ldots
\]
where dots stand for  divergent term which has no imaginary 
part and does not contribute to the spectral density in zero width
approximation.
A set of finite integrals appearing in eq.~(\ref{dG1b})
can be computed by the technique which has been already 
applied for the analysis
of the correction to the $l=0$ partial wave.
The results for relevant integrals read
\[
\int_0^\infty e^{-z}L^3_n(z)L^3_m(z)\ln(z)z^3dz=
\left\{
\begin{array}{c}
(m+1)(m+2)(m+3)\Psi_1(m+4), \quad m=n \\
{\displaystyle -{(n+1)(n+2)(n+3)\over m-n}}, \quad m>n \\
\end{array}
\right.
\]
\[
J_0(m)=\int_0^\infty e^{-z}L^3_m(z)\ln(z)dz=
-2\Psi_1(m+1)-4\gamma_E +3,
\]
\be
J_1(m)=\int_0^\infty e^{-z}L^3_m(z)\ln(z)zdz= 
(m+1)(-\Psi_1(m+2)-2\gamma_E +2),
\label{integ3}
\ee
\[
J_2(m)=\int_0^\infty e^{-z}L^3_m(z)\ln(z)z^2dz=
{(m+1)(m+2)\over 2}\left(-\Psi_1(m+3)-2\gamma_E +{3\over2}\right) .
\]
Substituting these formulae in eq.~(\ref{dG1c})
we arrive at the final result for the correction to the 
$l=1$ partial wave at the origin
\[
\Delta_1G_1(0,0,k)
={\al_s\beta_0\over 2\pi}{\lambda m_tk^2\over 18\pi}\left(
\sum_{m=0}^\infty \tilde F(m)^2(m+1)(m+2)(m+3)
\left(L(k)+\Psi_1(m+4)\right)
\right.
\]
\[
-2\sum_{m=1}^\infty\sum_{n=0}^{m-1}
\tilde F(m)\tilde F(n){(n+1)(n+2)(n+3)\over m-n} +2\sum_{m=0}^\infty 
\tilde F(m)\bigg(2J_0(m)+(m+1)(m+2)L(k)
\]
\be
\left.
+(1+\nu)(J_1(m)+(m+1)L(k))+{\nu(\nu+1)\over 2}(J_2(m)+2L(k))\bigg)
+I(k) \right).
\label{dG1}
\ee
Note that it is quite straightforward to extend our method to
calculation of the correction to a component 
of the Green function with arbitrary $l$. 

The Green function in every order of the nonrelativistic
perturbation theory has to  be written in the  
form which includes 
only single  poles in the energy variable and is more natural
for the Green function  of a nonrelativistic Schr{\"o}dinger equation.
For the  Green function at the origin this representation reads 
\be
G(0,0,E)=\sum_{m=0}^\infty{|\psi_m(0)|^2\over E^{(0)}_m-E}+
{1\over \pi}\int_0^\infty{|\psi_{E'}(0)|^2\over E'-E}
dE'
\label{endenom}
\ee
where  $\psi_{m,E'}(0)$ is the wave function at the origin, 
$E^{(0)}_m$ is a $l=0$ bound state energy,
the sum goes over bound states 
and the integration in the second term 
is performed over the state of continuous part of the
spectrum.
In this way the corrections to Green function stemming from
the discrete part of the spectrum 
can be reduced to the 
correction to the Coulomb bound state energy levels
\be
E^{(0)}_m=-{\lambda^2\over m_t(m+1)^2}\left(1+
{\alpha_s\over 4\pi}2C_1^1\left[L(m)+\Psi_1(m+2)\right]\right)
\label{ecorr}
\ee
and to the values of Coulomb bound state wave functions at the origin 
\[
|\psi_m(0)|^2={\lambda^3\over \pi(m+1)^3}\bigg(1+
{\alpha_s\over 4\pi}C_1^1\left(
3L(m)-1-2\gamma_E+{2\over m+1}+\Psi_1(m+2)\right.
\]
\be
-2(m+1)\Psi_2(m+1)\bigg)\bigg)
\label{pscorr}
\ee
where 
\[
L(m)=\ln\left({(m+1)\mu_1\over C_F\al_sm_b}\right).
\]
For the derivative of the Green function at the origin
one has 
\be
\partial_{\bf x}\partial_{\bf y}\Delta_1G({\bf x},{\bf y},E)|_{x,y=0}=
\sum_{m=0}^\infty{|\psi'_m(0)|^2\over E^{(1)}_m-E-i0}+
{1\over \pi}\int_0^\infty{|\psi'_{E'}(0)|^2\over E'-E-i0}
dE'
\label{dendenom}
\ee
where  
\[
|\psi'_{m,E'}(0)|^2=
\partial_{\bf x}\psi^*_{m,E'}({\bf x})
\partial_{\bf y}\psi_{m,E'}({\bf y})|_{x,y=0}
\]  
and $E^{(1)}_m$ is a $l=1$ bound state energy. 
In NLO approximation these quantities read 
\be
E^{(1)}_m=-{\lambda^2\over m_t(m+2)^2}\left(1+
{\alpha_s\over 4\pi}
2C_1^1\left[L(m+1)+\Psi_1(m+4)\right]\right)\ ,
\label{decorr}
\ee
and
\[
|\psi'_m(0)|^2={\lambda^5\over \pi}{(m+1)(m+3)\over (m+2)^5}
\left(1+
{\alpha_s\over 4\pi}C_1^1\left(-{\pi^2\over 3}(m+2)-1+
5L(m+1)+5\Psi_1(m+4) 
\right.\right.
\]
\be
\left.\left.
+2\sum^{m-1}_{n=0}{(n+1)(n+2)(n+3)\over
(m+1)(m+3)(m-n)^2}\right)\right)\ .
\label{dpscorr}
\ee
The continuum contributions  in eqs.~(\ref{endenom},~\ref{dendenom}) 
can be directly found by subtracting the   
discrete part of these equations expanded around  the Coulomb 
approximation up to NLO from the  result obtained within the 
nonrelativistic perturbation theory for the Green function
and its derivative at the origin.  

\section{Effects of the finite width of top quark.}
As has been already mentioned only sufficiently large
$t$-quark decay width suppresses the nonperturbative  
effects of strong interactions at large $(\sim 1/\Lambda_{QCD})$
distances and makes the perturbation theory
applicable for the description of the $t$-quark threshold dynamics.
The accurate description is achieved in the framework of NRQCD
that provides the adequate leading order approximation for the
problem.
However main features of the physical situation 
can be reflected in a simpler approach to which we restrict ourselves
in the present paper.
According to the prescription of ref.~\cite{FK} the non-zero width 
can be taken into account by direct replacing $E\rightarrow E+i\G_t$
in the Green function near the production threshold. 
This is justified because the threshold 
dynamics is nonrelativistic one
and is rather insensitive to
the hard momentum details of $t$-quark decays. 
This prescription is not equivalent to the full 
NRQCD description in higher orders of perturbation theory 
in the strong coupling constant.
Also, strictly speaking this 
prescription is valid in its simple form 
only when the complete factorization
of the hard and soft contributions to the cross section takes place.
This is realized in the case of $R^{++}$ cross section. 
In the case of $R^{+-}$ cross section, however,
the nonrelativistic approximation is not able  
to describe properly the entire effect of the non-zero top quark
width \cite{ttgg}.
Indeed, eq.~(\ref{dG0}) for a nonvanishing width
in the limit $x\rightarrow 0$ has the divergent imaginary part
with the leading power singularity $\sim\G_t/x$ related to the
free Green function singularity and 
logarithmic singularity $\sim\G_t\al_s\ln x$
produced by the one Coulomb photon exchange. 
This clearly indicates that the coefficient in front 
of the constant term linear in $\G_t$ in the cross section 
gets a contribution from large momentum region and cannot
be obtained within the nonrelativistic theory. 
This constant can be obtained within the full 
relativistic theory. 
The correct treatment of the problem consists in matching NRQCD expression 
for this part of the cross section to 
the one obtained in full theory by direct inserting the (complex)
mass operator into the $t$-quark propagator. In the leading order in $\al_s$
this has been done semiphenomenologically in refs.~\cite{ttgg,BFK}.
In the present paper we limit ourselves to formulas obtained in~\cite{ttgg,BFK}
that are sufficient from phenomenological point of view. 
The contribution linear in the top quark width reads    
\[
R_{\G_t}^{+-}=64\pi Q_t^4N_c\int {{\bf p}^2\over m_t^2} d\rho 
\]
\be
={8Q_t^4N_c\G_t\over m_t}\int_{(m_W+m_b)^2}^{m_t^2}
{\left((9m_t^2-p^2)(m_t^2-p^2)\right)^{3/2}
\over(m_t^2-p^2)^2+4m_t^2\G_t(p^2)^2}{\G_t(p^2)dp^2\over 32\pi\G_tm_t^4}
=0.185{8Q_t^4N_c\G_t\over m_t}
\label{relg}
\ee
where ${\bf p}^2$ is the  Born amplitude squared, 
$d\rho$ is the relativistic phase space of two unstable particles
for $E=0$ (see ref.~\cite{BFK}), 
$\G_t(p^2)$ is the imaginary part of the $t$-quark mass operator
($\G_t(m_t^2)=\G_t$) and the leading $t\rightarrow Wb$ decay channel
is taken into account. In eq.~(\ref{relg}) $m_W=80.3~{\rm GeV}$ 
is the $W$-boson mass \cite{PDG} and $m_b=4.8~{\rm GeV}$  is the 
$b$-quark pole mass \cite{PP1}. In the final expression for the 
cross section this contribution has to be multiplied by the 
hard normalization coefficient $C_h^{+-}(\al_s)$. Numerically this 
contribution to the cross section is suppressed in comparison with
one of the regular pure nonrelativistic terms of eq.~(\ref{dG0}) 
which saturate the total result. 

To find the coefficient in  front of the  $\G_t\al_s$ term in the 
imaginary part of eq.~(\ref{dG0}) or, equivalently, to fix the
parameter $\mu_f$ one has to know
the one-gluon correction to
Born approximation. 
This correction can be obtained from 
QED result \cite{TA} and for $E=0$ is equal to $\pi\lambda |{\bf p}|$.
A corresponding contribution to the cross section reads
\[
R_{\al_s\G_t}^{+-}=64\pi^2 Q_t^4N_c
\int {\lambda|{\bf p}|\over m_t^2} d\rho 
\]
\be
={8Q_t^4N_c\lambda\G_t\over m_t^2}\int_{(m_W+m_b)^2}^{m_t^2}
{(9m_t^2-p^2)(m_t^2-p^2)
\over(m_t^2-p^2)^2+4m_t^2\G_t(p^2)^2}{\G_t(p^2)dp^2\over 8\G_tm_t^2}
={8Q_t^4N_c\lambda\G_t\over m_t^2}
\left(\ln{\left(m_t\over\G_t\right)}-2.07\right)
\label{relga}
\ee
where the term with the logarithmic dependence on $\G_t$
is separated and computed  analytically within the approximation 
$\G_t(p^2)=\G_t$. 
Following the general line of the effective 
field theory approach one has  to equate
the contribution to the cross section
determined by the   $O(\G_t\al_s)$ terms in the 
imaginary part of eq.~(\ref{dG0}) at $E=0$
to the relativistic expression~(\ref{relga}).
This equation results in the following matching relation 
\[
-\ln\left({m_t\G_t\over \mu_f^2}\right)
-2\gamma_E+{19\over 6}=\ln\left({m_t\over\G_t}\right)-2.07
\]
which fixes the auxiliary parameter $\mu_f=0.13 m_t$.   
Note that in this relation the logarithmic dependence of  eq.~(\ref{dG0})
on $\G_t$ exactly matches one of eq.~(\ref{relga})
so $\mu_f$ is related only to the hard scale of the
process as one expects from the general ground. 

The above analysis cannot be considered as a 
completely relativistic one -- Born amplitude and the 
Coulomb one-gluon correction are taken
in the leading order of the nonrelativistic expansion 
though the relativistic phase volume is used.  
This, however, is justified because the integrals
in eqs.~(\ref{relg},~\ref{relga}) are saturated within the 
region $|{\bf p}|\sim \sqrt{m_t\G_t}<<m_t$ $(p^2\sim m_t^2)$
where the nonrelativistic approximation works well. 
By the same reason the function  $\G_t(p^2)$
in these equations can be well approximated by its constant value $\G_t$.

Note that for non-vanishing width some of the singular 
terms in  eq.~(\ref{Dinteg}) which describes the
correction to the Green function also have imaginary part and 
contribute to the cross section. This contribution, however,
has very weak energy dependence and is suppressed
by an extra power of $\al_s$. Thus, this is beyond the 
accuracy of our approximation. 

\section{Resonances of the cross section 
in the next-to-next-to-leading order.}
Though the complete NNLO analysis of the cross section
is not still available, some important parameters of 
the cross section can be found in NNLO
approximation. 
Indeed, the position of the poles of the 
nonrelativistic Green function which correspond
to resonances in the cross section can be determined
within the nonrelativistic approximation~\cite{PP,MY,PP1,PY}. 
This has been explicitly done
for the spin triplet states in refs.~\cite{MY,PP1,PY}.
This analysis can be directly generalized to the spin singlet
states which are produced in the two photon collisions. The only 
difference is in the correction induced 
by the part of the Breit potential responsible for the 
hyperfine splitting. We, however, do not need  the complete
result because the most of 
resonances are smoothed out by the relatively large top quark width.
The numerical analysis shows that only the ground state
resonance in $l=0$ partial wave cross section 
$R^{++}(E)$ is distinguishable.
Its separation from others is not completely covered
by the infrared cutoff provided by the top quark width.
Indeed, using the pure Coulomb formulas for estimates within the order of
magnitude we find
\[
|E^{(0)}_0-E^{(0)}_1|={3\lambda^2\over 4m_t}\approx 0.6~{\rm GeV}
\]
to be compared with the top quark width $\G_t=1.43~{\rm GeV}$.
The spacing between next radial excitations for $l=0$ partial wave 
(which is also equal to the first one for $l=1$ partial wave) is much smaller
\[
|E^{(0)}_1-E^{(0)}_2|=
|E^{(1)}_0-E^{(1)}_1|=\frac{5}{36}{\lambda^2\over m_t}\approx
0.11~{\rm GeV}
\]
and is completely smeared out with the top quark width.
Therefore it is numerically justified to treat 
separately only the first resonance in  
$l=0$ partial wave cross section 
and to sum all corrections to the denominator
of the Green function to avoid the appearance
of the double pole in the correction which is large at the resonance 
energy. For all other states this procedure makes no numerical
difference because the infrared cutoff is sufficiently large 
to smear out all picks in the cross section.
For $R^{+-}(E)$ cross section the lowest contributing state 
has $l=1$ and is rather close to other states and therefore 
is indistinguishable as a separate contribution.
   
Using the results of refs.~\cite{PP,MY,PP1,PY}
it is straightforward to find the energy of this resonance 
in NNLO approximation counted from the threshold 
\[
E_{res}(\gamma\gamma\rightarrow t\bar t)\equiv E^{(0)}_0
=-{\lambda^2\over m_t}\left(1+{\alpha_s\over 4\pi}
2C_1^1\left(L(\lambda)+1-\gamma_E\right)
+{\alpha_s\over 4\pi}^2\bigg(2C_1^2(L_1(\lambda)+1-\gamma_E)
\right.
\]
\[
\left.
+(C_1^1)^2\left((L(\lambda)-\gamma_E)^2+1-{\pi^2\over 3}-\Psi_3(1)\right)
+2C_2^2\left((L_2(\lambda)+1-\gamma_E)^2-1+{\pi^2\over 6}\right)
\right)
\]
\be
\left.
+C_F^2\al_s^2\left({C_A\over C_F}+{21\over 16}\right)
\right).
\label{e1}
\ee
where  
\[
L_1(\lambda)=\ln\left({\mu_se^{C_0^2/C_1^2}\over 2\lambda }\right),
\qquad L_2(\lambda)=\ln\left({\mu_s\over 2\lambda}\right),
\]
\[
C_0^2=\left({\pi^2\over 3}+4\gamma_E^2\right)\beta_0^2
+2(\beta_1+2\beta_0a_1)\gamma_E+a_2,
\]
\[
C_1^2=2(\beta_1+2\beta_0a_1)+8\beta_0^2\gamma_E,
\qquad 
C_2^2=4\beta_0^2,
\]
\[
a_2= \left({4343\over 162}+6\pi^2-{\pi^4\over 4}
+{22\over3}\zeta(3)\right)C_A^2-
\left({1798\over 81} + {56\over 3}\zeta(3)\right)C_AT_Fn_f
\]
$$
-\left({55\over 3} - 16\zeta(3)\right)C_FT_Fn_f
+\left({20\over 9}T_Fn_f\right)^2,
$$
\[
\beta_1={34\over 3}C_A^2-{20\over 3}C_AT_Fn_f-4C_FT_Fn_f
\]
and $\zeta(z)$ is the Riemann $\zeta$-function. 
This value is related to the energy of the bound state 
resonance of the top quark production in $e^+e^-$ annihilation
by the  hyperfine splitting
\be
E_{res}(e^+e^-\rightarrow t\bar t)-
E_{res}(\gamma\gamma\rightarrow t\bar t)
={4\over 3}
{\lambda^2\over m_t}C_F^2\al_s^2\ .
\label{de}
\ee

This NNLO result provides us with the 
information on the convergence of the perturbative expansion
and, in particular, restricts the allowed region of the 
normalization point which can be used for reliable 
estimates. In fact, the normalization points of $\al_s$
entering the coefficients $C_h$ and the nonrelativistic Green function
can be different when NNLO corrections are considered. The difference
between the normalization points of the hard and soft  
corrections can be noticed only in higher orders 
of perturbative expansion. This gives an additional 
possibility to improve the convergence
of the perturbation theory. The typical hard scale
of the problem is the heavy quark mass $m_t$. 
The soft physical scale of the problem 
is determined by the natural infrared cutoff related to the top quark width
$\sqrt{m_t\G_t}$ that measures the distance to the nearest 
singularity in the complex energy plane and/or by the characteristic scale of the 
Coulomb problem $\lambda$ {\it i.e.} $\mu_s\sim 15~{\rm GeV}$. 
For top quark both scales are rather close to each other that makes
possible
a uniform description both perturbative QCD and Coulomb resonance effects.
Indeed,
for 
$\mu_s\sim 15~{\rm GeV}$
the NLO correction to the resonance energy 
reaches its minimal magnitude (see Fig.~1). However,
from Fig.~1 we see that at this scale the NNLO correction is 
large and the series for the resonance energy diverges.
At first glance this seems to contradict our physical intuition.
However, since the normalization scale is defined in a rather 
artificial $\overline {\rm MS}$ scheme that has little to do with
peculiarities of 
$t\bar t$ physics being originally designed for describing
a massless quark approximation in $e^+e^-$ annihilation,
there is no reason for a literal coincidence of $\mu_s$
parameter with any physical scale of the process.
The relative weight of the NNLO corrections is stabilized at 
$\mu_s\sim 40~{\rm GeV}$ where the series for the resonance energy
looks like $E_{res}=E^{LO}_{res}(1+0.6+0.5)$. Moreover, for the larger 
values of the normalization point
the $\mu$ dependence of the cross section becomes small
(see the numerical analysis of the next Section).
Thus our analysis implies the optimal choice of normalization point 
in the range $\mu_s$\raisebox{-3pt}{$\stackrel{>}{\sim}$}$40~{\rm GeV}$.
The convergence of the series, however, is not fast
and the uncertainty related to the truncation of the series
exceeds those due to the nonperturbative effects
which are suppressed parametrically as $(\Lambda_{QCD}/\lambda)^4$
\cite{VL}. 
This implies relatively large NNLO corrections
to the $R^{++}$ cross section (similar effect takes place 
in $e^+e^-\rightarrow t\bar t$  
top quark threshold production \cite{Hoang,Mel}). 
Note that the apparent convergence for the resonance energy can be
slightly improved by changing the definition of the top quark mass and 
using $\overline {\rm MS}$ scheme for defining the mass parameter of
the top quark instead of its (perturbative) pole mass the convergence 
of the series for the resonance energy becomes better. 

\section{Summary and conclusion.}
Thus in the present paper the  
total cross section of the top quark pair production near the
threshold in $\gamma\gamma$ collision is computed analytically 
up to the next-to-leading order in perturbative and nonrelativistic
expansion for general photon helicity with the top quark
width being taken into account. 

To demonstrate the significance of the NLO corrections
we plot the functions  $R^{++}(E)$  and $R^{+-}(E)$ in Born $(\al_s=0)$,
leading order\footnote{In the leading order  approximation $C_h^{++}=C_h^{+-}=1$
and the Coulomb Green function is used.} 
and NLO approximations in Fig.~2 and Fig.~3. 
We use 
$m_t=175~{\rm GeV}$, $\G_t=1.43~{\rm GeV}$ and $\al_s(m_Z)=0.118$
as typical numerical values of corresponding parameters 
\cite{PDG}, $\mu_h=m_t$ for the hard normalization 
scale and $\mu_s=25~{\rm GeV}$, $50~{\rm GeV}$, and 
$75~{\rm GeV}$ for the soft normalization scale  
(in the leading order approximation the upper curves 
correspond to $\mu_s=25~{\rm GeV}$
and lower curves correspond to $\mu_s=75~{\rm GeV}$).
The results are almost independent of $\mu_h$
and their dependence on the soft normalization scale 
decreases with increasing $\mu_s$.  

The main results that we have obtained from the study 
of the NLO corrections to the 
total cross section of the top quark-antiquark 
pair production near 
threshold in $\gamma\gamma$ collision are:  
\begin{itemize}
\item The typical size of the NLO corrections is $20\%$ for
$R^{++}(E)$ and $10\%$ for $R^{+-}(E)$.
\item Inclusion of the NLO corrections leads to considerable
stabilization of the theoretical results for the cross sections
against changing the normalization point.
\item Relatively large top quark width smears out all
the resonances in $R^{+-}(E)$ and only the ground state
resonance survives in $R^{++}(E)$. 
The inclusion of the higher order corrections makes this peak
more distinguishable.
Thorough experimental study of the location and shape of this pick
can provide precise data for extraction of the mass and width 
of top quark
and the numerical value of the strong coupling constant.
\item The NNLO correction to the resonance energy in 
$\gamma\gamma\rightarrow t\bar t$ top quark threshold production
is relatively large that implies large NNLO corrections
to the $R^{++}$ cross section.
\end{itemize}

\vspace{10mm}
\noindent
{\large \bf Acknowledgments}\\[2mm]
We are thankful to Z.Merebashvili
for 
suggesting us to have a look at
the problem
during the Quarks-98 Seminar. This work is partially supported
by Volkswagen Foundation under contract
No.~I/73611. A.A.Pivovarov is supported in part by
the Russian Fund for Basic Research under contracts Nos.~96-01-01860
and 97-02-17065. The work of A.A.Penin is supported in part by
the Russian Fund for Basic Research under contract
97-02-17065.

\newpage

\section*{Figure captions}

\noindent
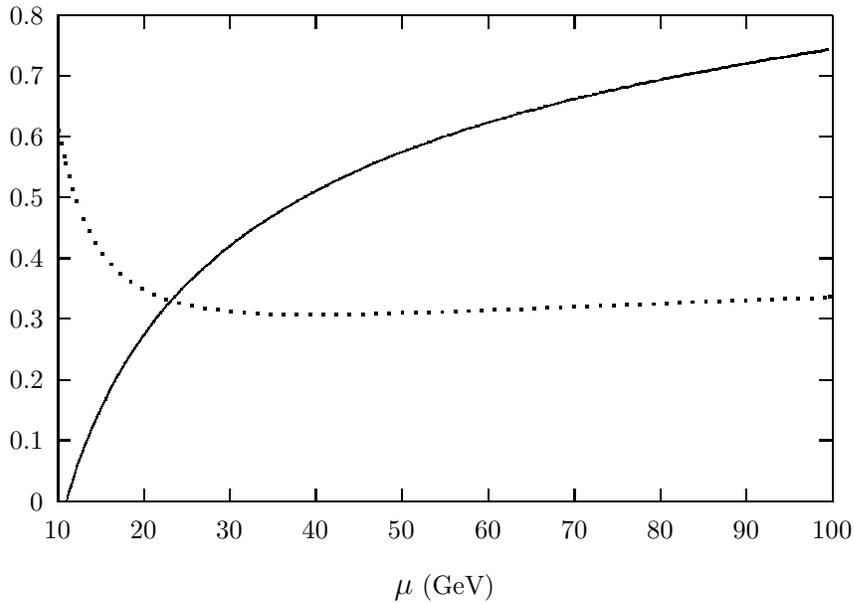
{\bf Fig. 1.} 
The relative weight of the 
NLO corrections $E_{res}^{NLO}/E_{res}^{LO}-1$  (solid line)
and NNLO corrections $E_{res}^{NNLO}/E_{res}^{NLO}-1$ (dotted line)
to the resonance energy  as a function of the  normalization 
point $\mu_s$.

\noindent
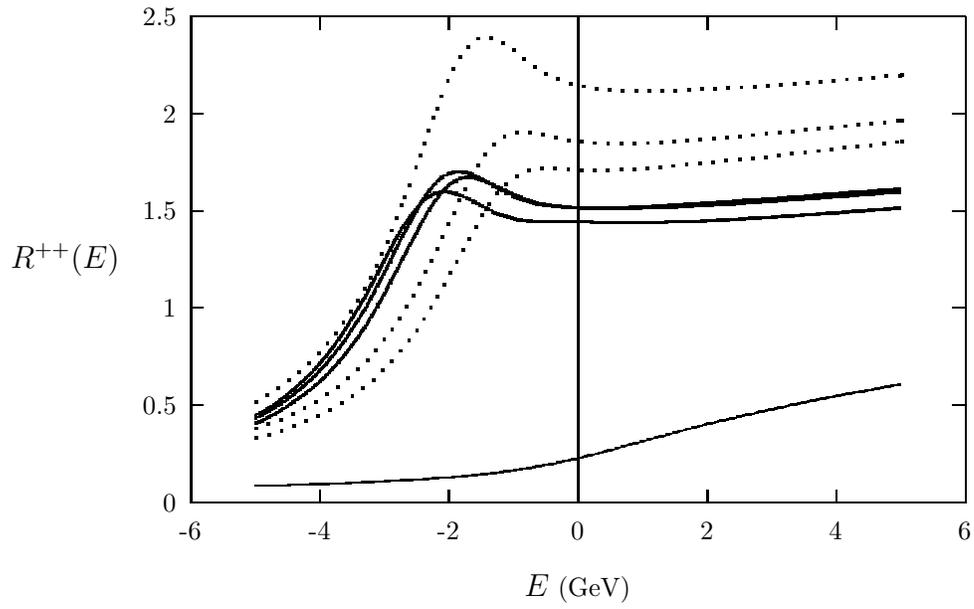
{\bf Fig. 2.} The normalized total cross section  $R^{++}(E)$
of top quark pair production for  the  colliding photons 
of the same  helicity in leading order
(dotted lines) and NLO (bold solid lines) for 
$\mu_s=25~{\rm GeV},~50~{\rm GeV}$ and $75~{\rm GeV}$.  
The normal solid line corresponds to the Born approximation.

\noindent
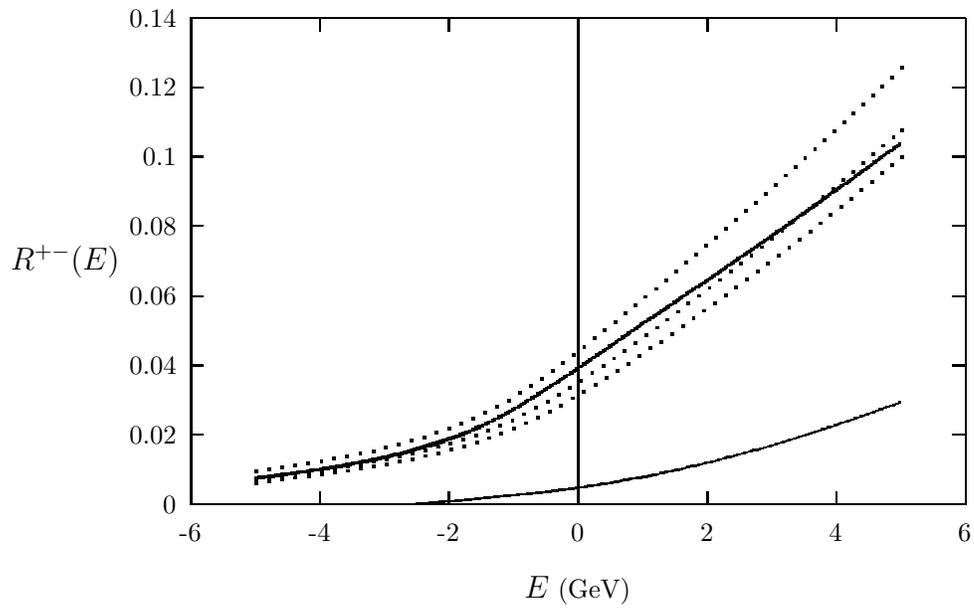
{\bf Fig. 3.} The normalized total cross section $R^{+-}(E)$
of top quark pair production for  the  colliding photons 
of the opposite  helicities in leading order
(dotted lines)  for $\mu_s=25~{\rm GeV},~50~{\rm GeV}$ and $75~{\rm GeV}$
and NLO (bold solid lines) for $\mu_s=50~{\rm GeV}$ 
(the NLO result  is given for only one normalization point because 
it has very weak dependence on the normalization scale).  
The normal solid line corresponds to the Born approximation.

\begin{center}
\vspace{5mm}

\setlength{\unitlength}{0.240900pt}
\ifx\plotpoint\undefined\newsavebox{\plotpoint}\fi
\sbox{\plotpoint}{\rule[-0.200pt]{0.400pt}{0.400pt}}%


\vspace{5mm}
{\bf Fig. 3.}

\end{center}

\end{document}